\documentclass[10pt,aps,twocolumn,showpacs,pra]{revtex4}
\usepackage{amssymb,amsfonts,amsmath,wrapfig,mathrsfs,mathbbol}
\usepackage{graphicx}

\newcommand{\fune}[1]{#1 ({\bf r},t)}
\newcommand{\bra}[1]{|#1 \rangle}

\begin{document}

\title{Measuring Two-Photon Orbital Angular Momentum Entanglement}

\author{G.F. Calvo, A. Pic\'{o}n and A. Bramon}
\affiliation{Grup de F\'{\i}sica Te\`{o}rica, Universitat Aut\`{o}noma de Barcelona, 08193 Bellaterra (Barcelona), Spain}

\date{\today}

\begin{abstract}
We put forward an approach to estimate the amount of bipartite spatial entanglement of down-converted photon states correlated in orbital angular momentum and the magnitude of the transverse (radial) wave vectors. Both degrees of freedom are properly considered in our framework which only requires azimuthal local linear optical transformations and mode selection analysis with two fiber detectors. The coincidence distributions predicted by our approach give an excellent fit to the distributions measured in a recent experiment aimed to show the very high-dimensional transverse entanglement of twin-photons from a down-conversion source. Our estimate for the Schmidt number is substantially lower but still confirms the presence of high-dimensional entanglement.
\end{abstract}
\pacs{03.67.Mn, 42.50.Dv, 42.65.Lm}

\maketitle

\section{INTRODUCTION}
\label{sect:intro}
The phenomenon of quantum entanglement whereby distant systems can manifest perfectly random albeit perfectly correlated behavior is now recognized as the essential ingredient to perform tasks which cannot be accomplished with classically correlated systems~\cite{OWQC}. The presence of entanglement has been traditionally revealed in the violation of Bell-type inequalities~\cite{Clauser}. However, detecting such a violation does not provide in general a measure of the amount of entanglement. This is particularly significant in systems correlated in multidimensional degrees of freedom~\cite{Collins}. Several techniques have been proposed to assess the presence of entanglement for different quantum scenarios. These include state tomography~\cite{White,Thew,Altepeter}, that yields a complete reconstruction of a quantum state but requires many setting measurements, entanglement witnesses~\cite{Witness}, which detect some entangled states with considerably less measurements, and the experimental determination of concurrence~\cite{Wootters,Walborn06}. In Ref.~\cite{Walborn06}, by using two copies of a down-converted two-photon state entangled in polarization and transverse momentum, the measurement of concurrence was achieved with a single, local measurement on one of the photons.    
\par
Although bipartite entanglement is well understood, finding experimentally feasible procedures to quantify it for systems correlated in multidimensional degrees of freedom turns out to be quite challenging and relevant. Indeed, access to higher dimensional Hilbert spaces in which information can be encoded and manipulated has recently attracted great interest, with proof-of-principle demonstrations using quantum communication protocols in three-level systems (qutrits), such as entanglement concentration~\cite{Vaziri03}, quantum bit commitment~\cite{Langford} and quantum coin-tossing~\cite{Gabi05}. Likewise, a complete characterization of states hyperentangled in polarization, orbital angular momentum and frequency has been experimentally implemented~\cite{Barreiro}. 
\par
The aim of this paper is to address the problem of how, by performing a certain set of local linear optical operations affecting one of two multidimensional spatial degrees of freedom in which two-photon states can be entangled, it is possible to obtain an explicit measure of the amount of bipartite transverse entanglement. Specifically, according to the interplay between both spatial degrees of freedom -orbital angular momentum (OAM) and the magnitude of the radial wave vectors- fundamentally different predictions are expected in the subsequent joint photodetection process, via mode-selection analysis with two fiber detectors preceded by azimuthal transformations (acting only on the OAM). The application of our framework is compared with the results of a recent experiment~\cite{Aiello05}.\par
The paper is organized as follows: Section II presents the general setting of the problem and includes the Schmidt decomposition technique for describing bipartite spatial entanglement. In Section III we reveal the generic features that arise in the photodetection coincidences according to the transverse structure of the two-photon wave function. Section IV illustrates the results found in Section III with an example of a realistic two-photon wave function that yields a full analytical solution to the problem. In Section V a closed-form expression for the Schmidt number is obtained in terms of easily accessible experimental parameters. We then proceed to exploit this Schmidt number in two interesting examples of optical transformations with azimuthal phase plates. These enable us to estimate the amount of spatial entanglement of a two-photon source. Conclusions of the paper are drawn in Section VI. Details of our calculations, together with some useful background material, have been included in two Appendices.
\par
\section{MODAL SCHMIDT DECOMPOSITION FOR TWO-PHOTON STATES}
\label{sec:SSD}
Two-photon pure quantum states are described in a Hilbert space by a continuous bilinear superposition of spatiotemporal multimode states. Sources of such nonclassical states of light are mostly realised in the process of spontaneous parametric down-conversion \cite{Klyshko}, where an intense quasimonochromatic laser pump illuminates a crystal endowed with a quadratic nonlinearity producing pairs of photons (idler and signal). Conservation of energy and momentum impose that the state be spectrally and spatially correlated. Here we explore entanglement involving spatial degrees of freedom that depend on the transverse structure of these states. For simplicity, we assume that the down-converted photons are linearly polarized, monochromatic and frequency degenerated. The two-photon state can then be written as $\vert\psi\rangle = \int \textrm{d}{\bf q}_{i}\textrm{d}{\bf q}_{s}\Phi({\bf q}_{i},{\bf q}_{s})\hat{a}^{\dagger}({\bf q}_{i})\hat{a}^{\dagger}({\bf q}_{s})\vert\textrm{vac}\rangle$, where ${\bf q}_{i,s}$ are the transverse components of the idler and signal wave vectors. Under conditions of paraxial and nearly collinear propagation of the pump, idler and signal photons, the two-photon amplitude $\Phi$ is given by~\cite{Downconversion} $\Phi({\bf q}_{i},{\bf q}_{s})=E({\bf q}_{i}+{\bf q}_{s})G({\bf q}_{i}-{\bf q}_{s})$. The function $E$ represents the transverse profile of the pump, whereas $G$ originates from the phase-matching condition in the longitudinal direction and depends on the specific orientation and cut of the nonlinear crystal. Since the arguments of $E$ and $G$ enforce correlations in different manifolds of the idler and signal wave vector space, it is the global structure of $\Phi$ the one dictating the entanglement degree of $\vert\psi\rangle$.
\par
In order to extract the amount of entanglement contained in $\Phi$, one may resort to the Schmidt decomposition that provides the spatial information modes of the two-photon pair. Suppose that the transverse spatial frequency field of the pump beam has a Gaussian profile of the form $E({\bf q}_{i}+{\bf q}_{s})\propto e^{-w_{0}^2\vert{\bf q}_{i}+{\bf q}_{s}\vert^{2}/4}$, where $w_{0}$ is the width (at the beam waist). The chosen pump profile peaks when its argument vanishes. This imposes that the idler and signal transverse wave vectors should be mostly anticorrelated  (${\bf q}_{i}\simeq-{\bf q}_{s}$). Remarkably, it was shown by Law and Eberly~\cite{Law04} that with such a Gaussian profile $E$ the normalised two-photon amplitude can be expressed as
\begin{eqnarray}
\Phi({\bf q}_{i},{\bf q}_{s}) = \sum_{\ell=-\infty}^{\infty} \sum_{n=0}^{\infty}(-1)^{\ell}\sqrt{\lambda_{\ell n}}\,u_{\ell,n}({\bf q}_{i})u_{-\ell,n}({\bf q}_{s})\, ,
\label{eq:Schmidt}
\end{eqnarray}
where $u_{\ell,n}({\bf q}_{i})=e^{i\ell\phi_{i}}v_{\ell,n}(q_{i})/\sqrt{2\pi}$ and $u_{-\ell,n}({\bf q}_{s})=e^{-i\ell\phi_{s}}v_{-\ell,n}(q_{s})/\sqrt{2\pi}$ are the normalised polar Schmidt mode functions for the idler and signal photons with topological charge $\ell$ and radial index $n$ corresponding to eigenvalues $\lambda_{\ell n}$ (they satisfy $1\geq\lambda_{\ell n}\geq0$ and $\sum_{\ell,n}\lambda_{\ell n}=1$) of the reduced density matrices for each photon. Knowledge of $\lambda_{\ell n}$ yields a direct measure of the degree of transverse entanglement given by the Schmidt number~\cite{Law00} $\mathcal{K}=(\sum_{\ell,n}\lambda_{\ell n}^{2})^{-1}$, which is the reciprocal of the purity of the idler and signal density matrices, it is invariant under free propagation and yields an average of the number of relevant spatial modes involved in the decomposition. The larger the value of $\mathcal{K}$, the higher the transverse entanglement. For instance, product states correspond to $\mathcal{K}=1$ (there is only one nonvanishing eigenvalue equal to $1$), whereas states with $\mathcal{K}>1$ are entangled. A distinguishing feature of decomposition~(\ref{eq:Schmidt}) is that it represents $\Phi$ in terms of a perfectly correlated discrete basis of paraxial eigenstates of the OAM operator along the direction of light propagation (with corresponding eigenvalues $\ell\hbar$)~\cite{OAMbook,CalvoPRA}, rather than in a continuous plane-wave modal expansion. The precise form of the radial idler and signal eigenmodes $v_{\ell,n}(q_{i})$ and $v_{-\ell,n}(q_{s})$ depends on the specific phase-matching function $G$. In particular, when $G$ is approximated by a constant there are striking consequences: Eq.~(\ref{eq:Schmidt}) becomes a (non-normalisable) superposition in which all OAM eigenstates have an equal weight and the radial dependence becomes just a global factor. It is important to emphasize that to attain decomposition (\ref{eq:Schmidt}) the appropriate choice of the widths, $w_{i}$ and $w_{s}$, for the idler and signal radial eigenmodes, has to be made. If the two-photon amplitude is of the form $\Phi({\bf q}_{i},{\bf q}_{s})=E({\bf q}_{i}+{\bf q}_{s})G({\bf q}_{i}-{\bf q}_{s})$, then $w_{i}=w_{s}\equiv w_{S}$, where $w_{S}$ is the so-called Schmidt width. For widths different from $w_{S}$ an additional summation over the radial indexes occurs, and the perfect correlation between the idler and signal radial modes is absent (see Appendix A). Moreover, the fact that the idler and signal mode functions are anticorrelated with respect to their topological charge numbers is a consequence of a more general process: the conservation of OAM, which is transferred from the pump photon (carrying zero OAM for a Gaussian mode) to the down-converted photon pair~\cite{Mair}. 
\par
\begin{figure}
\begin{center}
\hspace*{-0.15cm}
\hbox{\vbox{\vskip 0.0cm 
\includegraphics[width=60mm]{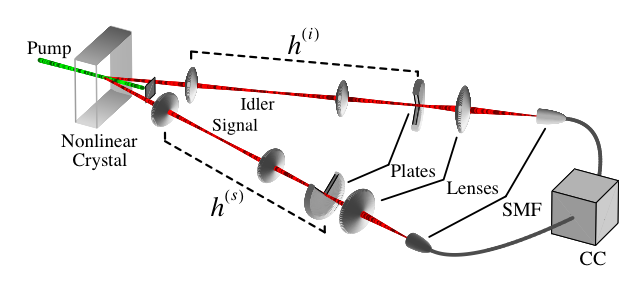}}}
\end{center}
\vspace*{-0.6cm}
\caption{\small (color online). Two-photon coincidence detection configuration. Down-converted idler and signal beams from a nonlinear crystal traverse two optical systems ($h^{(i,s)}$) which include azimuthal phase plates and are coupled into single-mode fibers (SMF) gated by a coincidence circuit (CC).}
\label{fig:Setup}
\end{figure}
\section{AZIMUTHAL TRANSFORMATIONS ON THE TWO-PHOTON STATE}
\label{sec:Azimuthal}
The usefulness of the Schmidt decomposition~(\ref{eq:Schmidt}) becomes apparent when analyzing the propagation of two-photon states through optical systems. Each of the intervening modes evolves and transforms independently of the others. It is also clear that the correlation properties displayed by the two-photon amplitude~(\ref{eq:Schmidt}) are preserved in the position representation. Therefore, suppose that the idler and signal photon beams, described now by the paraxial two-photon wave function in the transverse position representation~\cite{CalvoPRA,Bennink,Saleh} $\psi({\bf r}_{i},{\bf r}_{s}) =\langle {\bf r}_{i},{\bf r}_{s}\vert\psi\rangle= \sum_{\ell,n}(-1)^{\ell}\sqrt{\lambda_{\ell n}}u_{\ell,n}({\bf r}_{i})u_{-\ell,n}({\bf r}_{s})$, are each transmitted through different linear optical systems that include diffractive (or refractive) azimuthal phase plates (see Fig.~\ref{fig:Setup}). The role of the plates in each path is to imprint an azimuth-dependent phase factor on the incoming Schmidt modes. Their (separate) action on the two-photon wave function can be represented by the unitary and radially symmetric impulse response functions $h^{(i,s)}(\phi,\phi')$. These functions locally transform the spiral harmonic mode content of each photon via $e^{i\ell\phi}/\sqrt{2\pi}\to\sum_{\ell'}h_{\ell',\ell}^{(i,s)}e^{i\ell'\phi}/\sqrt{2\pi}$, with $h_{\ell',\ell}^{(i,s)}=\int_{0}^{2\pi}\!\int_{0}^{2\pi}\!\textrm{d}\phi\,\textrm{d}\phi'e^{i\ell\phi}e^{-i\ell'\phi'}h^{(i,s)}(\phi,\phi')/2\pi$, so that the resulting output two-photon wave function is 
\begin{eqnarray}
\psi_{out}({\bf r}_{i},{\bf r}_{s})\!&=&\! \sum_{\ell_{i},\ell_{s},\ell,n}(-1)^{\ell}\sqrt{\lambda_{\ell n}}\,h_{\ell_{i},\ell}^{(i)}\,h_{\ell_{s},-\ell}^{(s)}\nonumber\\
\!&\times&\!e^{i\ell_{i}\phi_{i}}v_{\ell,n}(r_{i})\,e^{i\ell_{s}\phi_{s}}v_{-\ell,n}(r_{s})/2\pi\, ,
\label{eq:Output}
\end{eqnarray}
where the initial perfect correlation in OAM is lost, remaining only that corresponding to the radial modes. 
\par
Upon traversing their respective linear optical systems the idler and signal photons, having OAM $\ell_{i}=\ell_{A}$, $\ell_{s}=\ell_{B}$ and radial indexes $n_{A,B}$, are detected in coincidence. By placing photodetectors at the output ports of suitable arrays of deterministic mode sorter interferometers~\cite{Leach02,Zambrini}, or computer-generated holograms~\cite{Vaziri03}, it is possible to distinguish modes bearing different OAM. In practice, the most straightforward procedure involves projecting into single-mode fibers, where the propagated mode has a fundamental Gaussian profile ($\ell_{A}=\ell_{B}=n_{A}=n_{B}=0$).
\par
The probability that the idler and signal photons will be projected into modes $u_{\ell_{A},n_{A}}$ and $u_{\ell_{B},n_{B}}$ is found to be $\mathcal{P}_{\ell_{A},\ell_{B}}^{n_{A},n_{B}}=\vert \int d{\bf r}_{i}d{\bf r}_{s}u_{\ell_{A},n_{A}}^{*}({\bf r}_{i})u_{\ell_{B},n_{B}}^{*}({\bf r}_{s})\psi_{out}({\bf r}_{i},{\bf r}_{s})\vert^{2}$, with an additional incoherent (and weighted) sum over the radial indexes $n_{A,B}$ when taking into account multi-mode detection. This photodection probability is expressed in terms of the spatial overlap of the Schmidt and the fiber modes {\em at the planes} where the phase plates are located. Notice that their corresponding widths, $w_{S}$ and $w_{G}$, are not necessarily equal, and thus the orthogonality between these modes when their radial indexes differ does not hold in general. To evaluate $\mathcal{P}_{\ell_{A},\ell_{B}}^{n_{A},n_{B}}$, we define $\mathcal{R}_{\ell}^{A,B}\equiv\sum_{n}\sqrt{\lambda_{\ell n}}\int r_{i}r_{s}dr_{i}dr_{s}v_{\ell_{A},n_{A}}^{(w_{G})}(r_{i})v_{\ell_{B},n_{B}}^{(w_{G})}(r_{s})$ $\times v_{\ell,n}^{(w_{S})}(r_{i})v_{-\ell,n}^{(w_{S})}(r_{s})$, where the two different widths of the radial modes have been specified for clarity. The coincidence probability can then be written as
\begin{eqnarray}
\mathcal{P}_{\ell_{A},\ell_{B}}^{n_{A},n_{B}}=\bigg\vert \sum_{\ell=-\infty}^{\infty}(-1)^{\ell}\mathcal{R}_{\ell}^{A,B}h_{\ell_{A},\ell}^{(i)}h_{\ell_{B},-\ell}^{(s)}\bigg\vert^{2}\, .\label{eq:Coincidence}
\end{eqnarray}
All the radial dependence is contained in the functions $\mathcal{R}_{\ell}^{A,B}$ that modulate the angular impulse response functions $h^{(i,s)}$. We emphasize that although the radial part of the Schmidt modes does not experience any significant transformation when the idler and signal beams traverse their respective linear optical systems, its proper inclusion in the detection process is essential. This is reflected in the structure of Eq.~(\ref{eq:Coincidence}) with the presence of $\ell$-dependent radial functions $\mathcal{R}_{\ell}^{A,B}$, and is a consequence of a non-constant phase-matching function $G$. Had $G$ been a constant then the input two-photon wave function~(\ref{eq:Schmidt}) would have been represented by a common radial function times a (non-normalisable) maximally entangled superposition of spiral harmonic modes. In this limiting case one derives a {\em fundamentally different} prediction for the coincidences: $\mathcal{P}_{\ell_{A},\ell_{B}}\propto\vert \sum_{\ell}(-1)^{\ell}h_{\ell_{A},\ell}^{(i)}h_{\ell_{B},-\ell}^{(s)}\vert^{2}$, where now all the radial dependence of the detected modes appears only as a global function.
\par
\section{AN EXACTLY SOLUBLE MODEL FOR THE TWO-PHOTON AMPLITUDE}
\label{sec:ExactModel}
To illustrate the previous results, suppose that the phase-matching function $G$ is also Gaussian (see Appendix C, where a justification of this model is provided), an approximation implying that photons are generated near the phase-matching region of wave vectors where the down-conversion process occurs most effectively. The normalised two-photon amplitude reads
\begin{eqnarray}
\Phi({\bf q}_{i},{\bf q}_{s}) = \frac{w_{0}b}{\pi}e^{-w_{0}^2\vert{\bf q}_{i}+{\bf q}_{s}\vert^{2}/4}e^{-b^{2}\vert{\bf q}_{i}-{\bf q}_{s}\vert^{2}/4}\, ,
\label{eq:TPA}
\end{eqnarray}
where $b<w_{0}$ plays the role of an {\em effective} width of the phase-matching function, and depends on the nonlinear crystal thickness, although no specific relation is assumed here. At variance with other models, where $G$ is often approximated by a constant ($b\to0$), the two-photon amplitude~(\ref{eq:TPA}) captures the relevant features of the transverse wave-vector correlation between the idler and signal photons, and provides an analytically amenable model that yields explicit formulas for Eq.~(\ref{eq:Coincidence}). We show in Appendix A that for the two-photon amplitude~(\ref{eq:TPA}), the Schmidt mode functions belong to the well-known Laguerre-Gaussian basis~\cite{Law04}, or, more generally, to a continuum family of spatial modes generated via metaplectic mappings (rotations on the orbital Poincar\'{e} sphere) from the Laguerre-Gaussian modes~\cite{CalvoPRA,Padgett99,CalvoOL}. Furthermore, we also find that the Schmidt eigenvalues are $\lambda_{\ell n}=(1-\xi^{2})^{2}\xi^{2\vert\ell\vert+4n}$, with $\xi=(w_{0}-b)/(w_{0}+b)$. This allows us to express the Schmidt number in the following closed form $\mathcal{K}=[(1+\xi^{2})/(1-\xi^{2})]^{2}$. Large values of $\mathcal{K}$ occur when $b\to 0$. The minimum $\mathcal{K}=1$ is attained when $b\to w_{0}$ (the two-photon amplitude~(\ref{eq:TPA}) becomes a separable function in the idler and signal wave vectors). The Schmidt width for any of the above families of eigenmodes is always the same: $w_{S}= \sqrt{2w_{0}b}$. We stress that the Schmidt width does not represent the actual cross-section widths $W_{i,s}$ of the idler and signal beams. These widths, that can be measured experimentally, can also be obtained by resorting to the partially reduced density matrix of the idler and signal photons. For amplitude~(\ref{eq:TPA}) one finds $W_{i,s}=\sqrt{2w_{0}^{2}+b^{2}}$, which is consistent with the widths employed in~\cite{Walborn05} in the thin crystal approximation ($b\to0$).
\par
Let us focus on the most usual encountered situation where the measured fiber modes are the fundamental Gaussian modes (of width $w_{G}$ at the phase plates). Since the involved eigenfunctions in the Schmidt decomposition~(\ref{eq:Schmidt}) have cylindrical symmetry, we use the Laguerre-Gaussian modes as the convenient computational basis to derive the radial functions $\mathcal{R}_{\ell}\equiv\mathcal{R}_{\ell}^{\ell_{A,B}=n_{A,B}=0}$ (see Appendix B). Remarkably, it turns out that the functions $\mathcal{R}_{\ell}$ can be cast in terms of a sole parameter $s$
\begin{eqnarray}
\mathcal{R}_{\ell}(s)=\frac{\Gamma^{2}(1+\frac{\vert\ell\vert}{2})}{\Gamma(1+\vert\ell\vert)}F\!\left(\frac{\vert\ell\vert}{2},\frac{\vert\ell\vert}{2};1+\vert\ell\vert;s^{2}\right)\!s^{\vert\ell\vert}\, ,\label{eq:LGCoincidence}
\end{eqnarray}
where $F(a,b;c;d)$ denotes the hypergeometric function, and
\begin{eqnarray}
s\!&\equiv&\! \frac{2\xi}{1+\xi^{2}+(1-\xi^{2})(w_{S}/w_{G})^{2}}\nonumber\\
\!&=&\!\frac{w_{0}^{2}-b^{2}}{w_{0}^{2}+b^{2}+(2w_{0}b/w_{G})^{2}}\, .\label{eq:s}
\end{eqnarray}
The parameter $s$ corrects $\xi$ by taking into account the spatial overlapping of the Schmidt and the fiber modes. Indeed, if and only if $w_{S}=w_{G}$, does $s=\xi$ hold. When $b\to0$ then $s\to1$, which corresponds to a constant phase-matching function. The functions $\mathcal{R}_{\ell}(s)$ increase monotonically from $\mathcal{R}_{\ell}(0)=0$ to $\mathcal{R}_{\ell}(1)=1$ for $\ell\neq 0$ ($\mathcal{R}_{0}(s)=1$). The fact that all $\mathcal{R}_{\ell}(s)$ depend on a single parameter $s$ will be exploited in the next Section to show how one can estimate the Schmidt number in a particular experimental scenario.
\par
\section{Experimental Schmidt Number}
\label{sec:ESN}
Let us now examine the problem of measuring the amount of transverse entanglement of two-photon sources which, in our case, is characterised by the Schmidt number $\mathcal{K}$. In principle, a complete quantum tomography of the two-photon state could yield the desired amount of entanglement~\cite{Langford}, but generally this would require a very large number of measurements, each for every possible pair of spatial modes, and this is technically demanding. Here we propose an alternative approach. In our simple model for the two-photon amplitude~(\ref{eq:TPA}) we have found that the radial part of the coincidence probabilities~(\ref{eq:Coincidence}) depends on a single parameter $s$. This parameter $s$ involves the characteristic widths appearing in the two-photon amplitude~(\ref{eq:TPA}), namely the pump beam width $w_{0}$ and the phase-matching width $b$, together with the fiber mode width $w_{G}$ (at the phase plate locations). The phase-matching width $b$ could be measured by scanning in the plane of detection, but in our case it is not necessary. If, instead, one rotates the azimuthal phase plates (maintaining the detectors fixed), then, the recording of the coincidence distributions allows one to extract the value of $s$ as a fitting parameter for $\mathcal{P}_{s}\equiv\mathcal{P}_{\ell_{A}=0,\ell_{B}=0}^{n_{A}=0,n_{B}=0}$ via Eqs.~(\ref{eq:Coincidence}) and (\ref{eq:LGCoincidence}). Therefore, if $s$ is conceived as a parameter to be directly measured (rather than $b$), the amount of transverse entanglement can now be written in the form
\begin{eqnarray}
\mathcal{K}=\frac{(1+2s\mu^{2})^{2}}{(1-s)(1+s+4s\mu^{2})}\, ,\label{eq:Schmidtnumber}
\end{eqnarray}
where $\mu\equiv w_{0}/w_{G}$. This {\em experimental} Schmidt number depends on quantities that are easily accessible: the fitting parameter $s$ and the widths $w_{0}$ and $w_{G}$ (the presence of the width ratio $\mu$ should be interpreted as a correcting geometrical factor). It increases from $\mathcal{K}=1$, when $s=0$, to infinity as $s\to1$. 
\par
By properly engineering the impulse response functions $h^{(i,s)}$ it is possible to enhance the sensitivity of $\mathcal{P}_{s}$ with $s$, thus improving the accuracy of the estimated $\mathcal{K}$. To this end, we consider two simple types of transparent azimuthal phase plates, and examine the dependence of Eq.~(\ref{eq:Coincidence}) when the idler and signal phase plates are mutually rotated a relative angle $\alpha\equiv\alpha_{i}-\alpha_{s}$. Their dispersionless impulse response functions are of the form $h^{(i,s)}(\phi,\phi')=e^{i\theta_{i,s}(\phi)}\delta(\phi-\phi')$. Owing to the initial perfect anticorrelation in OAM of the down-converted photon pairs, one expects that only when both photons are subjected to {\em complementary} azimuthal transformations the perfect anticorrelation is preserved and the coincidences are maximal. As soon as the phase plates are rotated in such a way that they are not longer oriented in a complementary arrangement, the photon coincidences decrease. Notice that the axes with respect of which the angles $\alpha_{i}$ and $\alpha_{s}$ are taken do not coincide. For symmetric phase plates these reference axes are inverted $180^{\circ}$. In what follows, we assume that each plate is characterised by a noninteger parameter $\eta=(n_{0}-1)d/\lambda$, where $d$ is the relative step height introduced by the plates, $n_{0}$ their refractive index, and $\lambda$ the wavelength of the idler and signal beams.
\par
\begin{figure}
\begin{center}
\hspace*{-0.3cm}
\hbox{\vbox{\vskip 0.0cm \includegraphics[width=60mm]{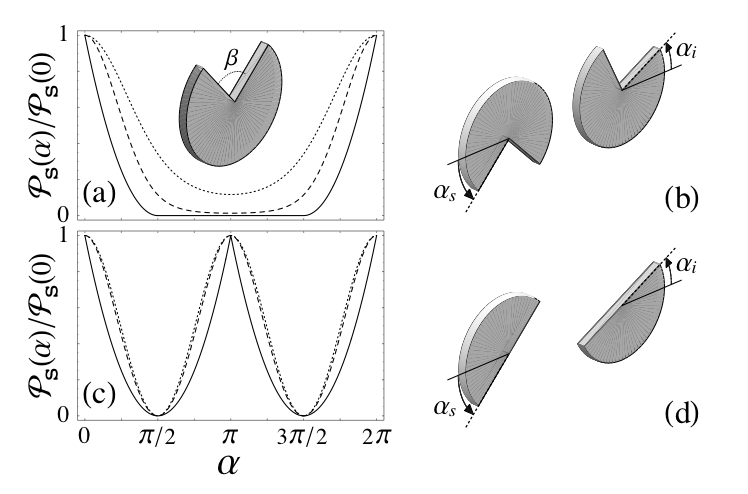}}}
\end{center}
\vspace*{-0.6cm}
\caption{\small (Left column) Normalised coincidence distributions as a function of the relative orientation between the idler and signal angular diaphragms. Solid, dashed and dotted curves are calculated for $s=1,0.7$, and $0.4$, respectively. (a) $\beta=\pi/2$ and (c) $\beta=\pi$, both with $\eta=0.5$. (Right column) The corresponding configurations of the angular diaphragms.} 
\label{fig:AD}
\end{figure}
The first phase plate type we consider is an angular diaphragm [see inset in Fig.~\ref{fig:AD}(a)]. It consists of a thin uniform dielectric circular slab with a ``cake-slice'' indentation that subtends an angle $\beta$, with a nonzero $\theta_{i}(\phi)=-\theta_{s}(\phi)\equiv 2\pi\eta$ (mod $2\pi$) only if $\alpha_{i,s}+\beta<\phi<\alpha_{i,s}+2\pi$. Similar angular diaphragms have been employed in proof-of-principle demonstrations of the uncertainty relation for angular position and OAM~\cite{Franke-Arnold}. The probability~(\ref{eq:Coincidence}) can be cast as $\mathcal{P}_{s}^{(AD)}(\alpha)=\left[\mathcal{R}_{0}(s)[(\beta-\pi)^2 + \pi^{2}\cot^{2}(\pi\eta)]\right.$ $\left.+8\sum_{\ell=1}^{\infty}\mathcal{R}_{\ell}(s)\cos(\ell\alpha)\sin^{2}(\ell\beta/2)/\ell^{2}\right]^{2}$. The main features of the normalized coincidence $\mathcal{P}_{s}^{(AD)}(\alpha)/\mathcal{P}_{s}^{(AD)}(0)$ are: (i) it does not depend on the integer part of $\eta$ (it suffices to consider $0<\eta<1$), the visibility being maximal when $\eta=1/2$; (ii) the coincidence distributions are identical whether the aperture angle is $\beta$ or $2\pi-\beta$ and symmetrical around $\alpha=\pi$; (iii) for $\beta\neq\pi$ and $\eta$ fixed, the visibility diminishes as $s$ decreases; and (iv) the maximum visibility always occurs in the limit $s\to1$ (constant phase-matching function and thus very high Schmidt number) where one has $\mathcal{P}_{s\to1}^{(AD)}(\alpha)=\pi^{2}\left[\pi[\cot^{2}(\pi\eta)-1] +\vert2\pi-\alpha-\beta\vert+\vert\alpha-\beta\vert\right]^{2}$. Figures \ref{fig:AD}(a) and (c) depict the characteristic profiles of the (normalised) $\mathcal{P}_{s}^{(AD)}$ when the aperture angles of the angular diaphragms are $\beta=\pi/2$ and $\beta=\pi$ [Figs. \ref{fig:AD}(b) and (d) show, respectively, the configurations of the angular diaphragms that yield the coincidences (a) and (c)]. Comparing Figs.~\ref{fig:AD}(a) and (c) one sees that the visibility of the former exhibits a much stronger variation with $s$ than the latter. This suggests that the phase plate configuration of Fig. \ref{fig:AD}(b) is preferred to that of Fig. \ref{fig:AD}(d) for achieving a more accurate estimation of the Schmidt number via Eq.~(\ref{eq:Schmidtnumber}). Finally, we should add that if $\eta$ is an integer number then the profiles of $\mathcal{P}_{s}^{(AD)}$ are constant (independent of $\alpha$).
\par
\begin{figure}
\begin{center}
\hspace*{-0.3cm}
\hbox{\vbox{\vskip 0.0cm \includegraphics[width=60mm]{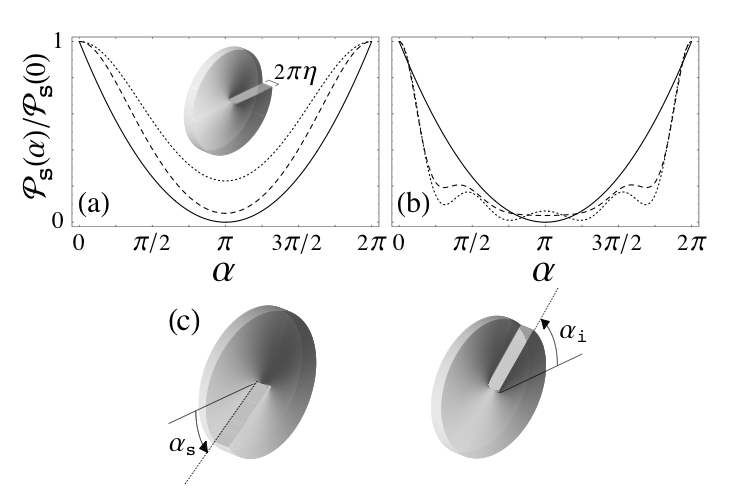}}}
\end{center}
\vspace*{-0.6cm}
\caption{\small Normalised coincidence distributions as a function of the relative orientation between the idler and signal spiral phase plates. Solid, dashed and dotted curves are calculated for $s=1,0.7$, and $0.4$, respectively. (a) $\eta=0.5$; and (b) $\eta=4.5$. (c) Configuration of the spiral phase plates.} 
\label{fig:SPP}
\end{figure}
The second type of azimuthal (refractive) component is a spiral phase plate [see inset in Fig.~\ref{fig:SPP}(a)]~\cite{OAMbook,Berry}. Its impulse response function includes the phase dependence $\theta_{i}(\phi)=-\theta_{s}(\phi)=\eta\phi$, for $\alpha_{i,s}<\phi<\alpha_{i,s}+2\pi$. In this case, the joint probability~(\ref{eq:Coincidence}) can be written as $\mathcal{P}_{s}^{(SPP)}(\alpha)=\left\vert \sum_{\ell}\mathcal{R}_{\ell}(s)e^{i\ell\alpha}(\ell+\eta)^{-2}\right\vert^{2}$. At variance with $\mathcal{P}_{s}^{(AD)}$, the coincidence distributions exhibit the development of interference ripples as $s$ decreases and $\eta$ increases [see Figs.~\ref{fig:SPP}(a) and (b)]. When $s\to1$ a parabolic profile is obtained $\mathcal{P}_{s\to1}^{(SPP)}(\alpha)=\pi^{2}\left[\pi^{2}\cot^{2}(\pi\eta)+(\alpha-\pi)^{2}\right]/\sin^{2}(\pi\eta)$. In this limit the coincidences become independent of the integer part of $\eta$, and, as with angular diaphragms, the maximum coincidence visibility is attained (when $\eta=1/2$). 
\par
We have considered other types of azimuthal phase plates and found the same parabolic dependence of the coincidences as $s\to1$. This is consistent with the abovementioned prediction that for very high transverse entanglement the joint probability becomes independent of the radial structure of the Schmidt modes. On the other hand, the absence of a vanishing minimum in the coincidence distributions is a signature of a finite amount of transverse entanglement. This effect gives rise to a contribution in the photocounts which always exists regardless of the detector efficiencies [see Fig.~\ref{fig:AD}(a) and Figs.~\ref{fig:SPP}(a),(b)]. It can however be diminished for certain phase plate configurations [see Fig.~\ref{fig:AD}(c)] and/or by employing a large $w_{G}$ ($w_{G}\gg \sqrt{2w_{0}b}$). 
\par
\begin{figure}
\begin{center}
\hspace*{-0.3cm}
\hbox{\vbox{\vskip 0.0cm \includegraphics[width=60mm]{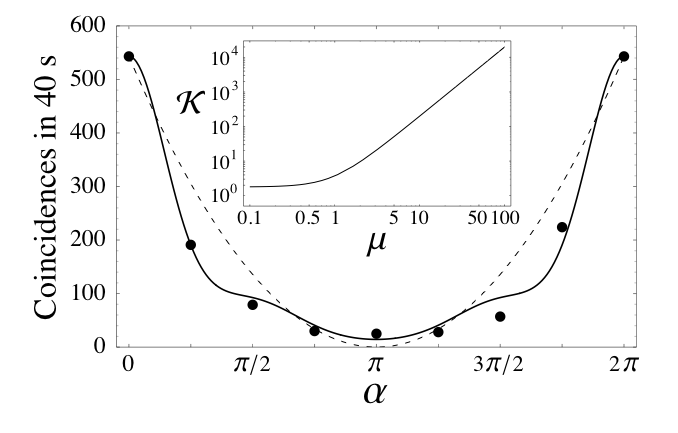}}}
\end{center}
\vspace*{-0.6cm}
\caption{\small Coincidence distribution dependence on the relative orientation between idler and signal spiral phase plates (SPP with $\eta=3.5$). Circles are experimental values from Ref.~\cite{Aiello05}, solid and dashed curves correspond to $\mathcal{P}_{s}^{(SPP)}$ fitted with $s=0.66$, and $\mathcal{P}_{s\to1}^{(SPP)}$, respectively. The slight asymmetry in the experimental values was probably caused by the presence of a small anomaly in the center of the spiral phase plates~\cite{Aiello05}. (Inset) Schmidt number $\mathcal{K}$ as a function of the width ratio $\mu=w_{0}/w_{G}$ for $s=0.66$.}
\label{fig:SPP2}
\end{figure}
Spiral phase plates were recently used in an elegant experiment aimed to show the high-dimensional spatial entanglement of a two-photon state from a down-conversion source~\cite{Aiello05}. The relevant parameters of the experiment are: $\eta=3.5$ for the phase plates, $\lambda=0.8$ $\mu$m, pump width $w_{0}=780$ $\mu$m, and thickness of the nonlinear crystal $L=1$ mm~\cite{Aiello05}. According to the model presented in Ref.~\cite{Aiello04}, which corresponds to our $\mathcal{P}_{s\to1}^{(SPP)}$, it was concluded that $\mathcal{K}>3700\pm100$. Figure~\ref{fig:SPP2} depicts the predicted distribution for $\mathcal{P}_{s}^{(SPP)}$ (dashed line) together with the measured coincidences reported in Ref.~\cite{Aiello05}. According to our approach, a probability distribution $\mathcal{P}_{s}^{(SPP)}$ fitted with $s=0.66$ (solid line in Fig.~\ref{fig:SPP2}) shows excellent agreement with the experimental results. Equation~(\ref{eq:Schmidtnumber}) then suggests that the corresponding Schmidt number should be much smaller than the above $\mathcal{K}$. However, we cannot perform a definite estimate for $\mathcal{K}$ since the value of the fiber mode width $w_{G}$ at the location of the spiral phase plates, needed to calculate $\mu=w_{0}/w_{G}$ appearing in Eq.~(\ref{eq:Schmidtnumber}), was not provided in Ref.~\cite{Aiello05}. The inset in Fig.~\ref{fig:SPP2} plots $\mathcal{K}$ for a wide range of ratios $\mu=w_{0}/w_{G}$. For instance, using our fitting parameter $s=0.66$, together with $w_{0}=780$ $\mu$m and assuming $w_{G}\gtrsim250$ $\mu$m, leads to $\mu\lesssim3$ and an estimate for $\mathcal{K}\lesssim20$, at least two orders of magnitude smaller than the quoted $\mathcal{K}$ in Ref.~\cite{Aiello05}. A very large value for $\mathcal{K}$ could only be expected if, in the experiment, the fiber mode width $w_{G}$ in the plane containing the phase plates was much smaller than the pump width $w_{0}$. 
\par
\section{Conclusions}

In view of the situation described above, it is reasonable to conclude that additional experimental results and further theoretical analyses -a recent example can be found in Ref.~\cite{Aiello06}- are necessary and of interest in order to clarify the problem of quantifying the amount of transverse entanglement of photon pairs produced in down-conversion sources. If future values for the Schmidt number coming from accurate and solid measurements confirm the estimate of Ref.~\cite{Aiello05} and cannot be reproduced within our approach, one should conclude that the conventional Gaussian profile constitutes a poor approximation for the phase-matching function $G$ in Eq.~(\ref{eq:TPA}). However, the ability of our predicted coincidence distributions to fit the experimental distributions of Ref.~\cite{Aiello05} suggests the correctness of our analysis and that the value for the Schmidt number quoted in Ref.~\cite{Aiello05} could be overestimated. In this case, our approach not only would provide a simple an accurate procedure to identify the relevant spatial modes and extract the degree of transverse entanglement; it would go in fact beyond by characterizing the action of all local bipartite azimuthal optical transformations on a broad family of two-photon states using only two detectors. Our results could also be of interest in other fields such as in identifying the intervening spatial modes when preparing well-controlled superpositions of photon states carrying OAM (to be coherently transferred onto Bose-Einstein condensates~\cite{Kapale,Andersen} or by entangling photons with ensembles of cold atoms~\cite{Inoue}), in high-resolution ghost diffraction experiments with thermal light~\cite{Ferri}, or in decoherence processes such as the entanglement sudden death mediated by the simultaneous action of several weak noise sources on bipartite photon systems~\cite{Eberly}.
\par
\acknowledgments
We thank S. Barnett, D. Diego, G. Molina-Terriza, M.J. Padgett and S. Walborn for useful discussions and gratefully acknowledge financial support from the Spanish Ministry of Science and Technology through Project FIS2005-01369, Juan de la Cierva Grant Program, CONSOLIDER2006-00019 Program and CIRIT Project SGR-00185.
\par
\appendix
\section{}
In this Appendix we outline the derivation of the eigenvalues and eigenfunctions in the Schmidt decomposition~(\ref{eq:Schmidt}) when the two-photon amplitude $\Phi$ is given by Eq.~(\ref{eq:TPA}). We start with some well-known facts.
\par
For a general two-photon pure state, $\vert\psi\rangle$, it is possible to express $\vert\psi\rangle$ as a bilinear sum of idler and signal basis states $\vert \tau_{i,s}\rangle$ belonging to the Hilbert space of the system 
\begin{eqnarray}
\vert \psi\rangle = \sum_{\tau_{i},\tau_{s}} C_{\tau_{i},\tau_{s}}\vert \tau_{i}\rangle\otimes\vert \tau_{s}\rangle\, ,
\label{eq:psi1}
\end{eqnarray}
where $\tau_{i}$ and $\tau_{s}$ label the set of quantum numbers for the idler and signal photons, respectively, whereas the coefficients $C_{\tau_{i},\tau_{s}}$ describe the probability amplitudes for each tensor product of basis states. Our first aim is to evaluate the coefficients $C_{\tau_{i},\tau_{s}}$ for the down-converted state $\vert\psi\rangle = \int \textrm{d}{\bf q}_{i}\textrm{d}{\bf q}_{s}\Phi({\bf q}_{i},{\bf q}_{s})\hat{a}^{\dagger}({\bf q}_{i})\hat{a}^{\dagger}({\bf q}_{s})\vert\textrm{vac}\rangle$. Let $u_{\tau_{i,s}}({\bf q}_{i,s})=\langle\textrm{vac}\vert \hat{a}({\bf q}_{i,s})\vert\tau_{i,s}\rangle$ denote the basis wave functions in the transverse momentum representation. Inserting the closure relations $\sum_{\tau_{i,s}}\vert \tau_{i,s}\rangle\langle\tau_{i,s}\vert=\hat{\mathbb{1}}$, it can be readily seen that the coefficients $C_{\tau_{i},\tau_{s}}$ are given by 
\begin{eqnarray}
C_{\tau_{i},\tau_{s}}=\int\textrm{d}{\bf q}_{i}\textrm{d}{\bf q}_{s}\Phi({\bf q}_{i},{\bf q}_{s})u_{\tau_{i}}^{*}({\bf q}_{i})u_{\tau_{s}}^{*}({\bf q}_{s}).\label{eq:Ctau}
\end{eqnarray}
\par
Since we are interested in elucidating the correlation properties of the spatial degrees of freedom (OAM and the magnitude of the transverse radial wave vectors), we choose as the computational basis the complete set of normalised Laguerre-Gaussian modes~\cite{CalvoPRA}
\begin{eqnarray}
u_{\ell,n}(q,\phi)\!&=&\!\sqrt{\frac{w^{2}n!}{2\pi(\vert\ell\vert + n)!}}\left(\frac{wq}{\sqrt{2}}\right)^{\!\vert\ell\vert}\!L_{n}^{\vert\ell\vert}\!\left(\frac{w^{2}q^{2}}{2}\right)\nonumber\\ 
\!&\times &\!\exp\left(-\frac{w^{2}q^{2}}{4}\right)\exp\left[i\ell\phi-i\frac{\pi}{2}\left( 2n+\vert\ell\vert\right)\right]\nonumber\\ 
\!&\equiv&\! \exp\left(i\ell\phi\right)v_{\ell,n}(q)/\sqrt{2\pi} ,\label{eq:FOURIERLG}
\end{eqnarray}
where $q$ and $\phi$ denote the radial and azimuthal variables in momentum space, $w$ is the mode width (at the beam waist), and $L_{n}^{\vert\ell\vert}(x)$ are the associated Laguerre polynomials. The indices $\ell=0,\pm 1,\pm 2,\ldots$ and $n=0, 1, 2,\ldots$ represent the winding (or topological charge) and the number of nonaxial radial nodes of the modes. 
\par
Combining Eqs.~(\ref{eq:TPA}), (\ref{eq:Ctau}) and (\ref{eq:FOURIERLG}), it follows that
\begin{eqnarray}
C_{\ell_{i},\ell_{s}}^{n_{i},n_{s}}\!&=&\!\frac{w_{0}b}{\pi}\!\int \textrm{d}{\bf q}_{i}\,\textrm{d}{\bf q}_{s}\,e^{-w_{0}^2\vert{\bf q}_{i}+{\bf q}_{s}\vert^{2}/4}e^{-b^{2}\vert{\bf q}_{i}-{\bf q}_{s}\vert^{2}/4}\nonumber\\
\!&\times&\! e^{-i\ell_{i}\phi_{i}}e^{-i\ell_{s}\phi_{s}}v_{\ell_{i},n_{i}}(q_{i})\,v_{\ell_{s},n_{s}}(q_{s})\, .\label{eq:Cln}
\end{eqnarray}
By means of the well-known Anger-Jacobi identity $e^{-x\cos(\phi_{i}-\phi_{s})}=\sum_{m=-\infty}^{\infty}(-1)^{m}I_{m}(x)e^{im(\phi_{i}-\phi_{s})}$, where $I_{m}(x)$ is the modified Bessel function of the first kind, the two angular integrals yield the selection rule $m=\ell_{i}=-\ell_{s}\equiv\ell$. This shows that the idler and signal photons are perfectly anticorrelated with respect to their topological charge, which is a manifestation of OAM entanglement. Hence, $C_{\ell_{i},\ell_{s}}^{n_{i},n_{s}}=C_{\ell,-\ell}^{n_{i},n_{s}}\delta_{\ell\ell_{i}}\delta_{-\ell\ell_{s}}$, where
\begin{eqnarray}
C_{\ell,-\ell}^{n_{i},n_{s}}\!\!&=&\!\!(-1)^{\ell}2w_{0}b\!\int_{0}^{\infty}\!\!\!\!\int_{0}^{\infty}\!q_{i}\textrm{d}q_{i}q_{s}\textrm{d}q_{s}e^{-(w_{0}^2+b^{2})(q_{i}^{2}+q_{s}^{2})/4}\nonumber\\
\!&\times&\! v_{\ell,n_{i}}(q_{i})\,v_{-\ell,n_{s}}(q_{s})\,I_{\ell}\left(\frac{(w_{0}^{2}-b^{2})q_{i}q_{s}}{2}\right) .\label{eq:Cln2}
\end{eqnarray}
The first of the radial integrals in~(\ref{eq:Cln2}) can be performed by resorting to the following formula
\begin{eqnarray}
\int_{0}^{\infty} x^{\vert\ell\vert+1}L_{n}^{\vert\ell\vert}(x^{2})e^{-\gamma x^{2}} I_{\ell}(xy)\textrm{d}x\nonumber\\
=\frac{(\gamma-1)^{n}y^{\vert\ell\vert}}{2^{\vert\ell\vert+1}\gamma^{\vert\ell\vert+n+1}}e^{\frac{y^{2}}{4\gamma}}L_{n}^{\vert\ell\vert}\left(\frac{y^{2}}{4(\gamma-1)\gamma}\right)\!,\nonumber
\end{eqnarray}
valid for all real $y$, integers $n\geq0$ and $\ell$, and complex $\gamma$ ($\textrm{Re}(\gamma)>0$).
\par
The second radial integral can also be done analytically. However, a dramatic simplification occurs if the widths $w_{i}$ and $w_{s}$ of the idler and signal radial modes, which at this stage have not been specified, are properly selected: $w_{i}=w_{s}=\sqrt{2w_{0}b}$.  In this case, one obtains a second selection rule for the radial indices. Namely, $n_{i}=n_{s}\equiv n$. That is, with such a choice of the widths, which is unique, the idler and signal radial modes are perfectly correlated. But this shows that it is precisely for $w_{i}=w_{s}=\sqrt{2w_{0}b}$ when one derives the Schmidt decomposition. Thus, the Schmidt width, $w_{S}$, corresponds to $w_{S}=\sqrt{2w_{0}b}$. The coefficients reduce then to 
\begin{eqnarray}
C_{\ell_{i},\ell_{s}}^{n_{i},n_{s}}\!\!&=&\!\!(-1)^{\ell}\frac{4w_{0}b}{(w_{0}+b)^{2}}\left(\frac{w_{0}-b}{w_{0}+b}\right)^{\vert\ell\vert+2n}\nonumber\\
\!&\times&\!\delta_{\ell\ell_{i}}\delta_{-\ell\ell_{s}}\delta_{nn_{i}}\delta_{nn_{s}}.\label{eq:Cln3}
\end{eqnarray}
\par
Let $\xi=(w_{0}-b)/(w_{0}+b)$. It is now clear from Eq.~(\ref{eq:Cln3}) that the Schmidt eigenvalues are $\lambda_{\ell n}=(1-\xi^{2})^{2}\xi^{2\vert\ell\vert+4n}$. We therefore conclude that the Schmidt decomposition of the two-photon amplitude~(\ref{eq:TPA}) is 
\begin{eqnarray}
\Phi({\bf q}_{i},{\bf q}_{s})\!\!&=&\!\! (1-\xi^{2})\!\sum_{\ell=-\infty}^{\infty}\sum_{n=0}^{\infty}(-1)^{\ell}\xi^{\vert\ell\vert+2n}u_{\ell,n}({\bf q}_{i})\nonumber\\
&\times& u_{-\ell,n}({\bf q}_{s}), \label{eq:SchmidtPhi}
\end{eqnarray}
where $u_{\ell,n}({\bf q}_{i})=e^{i\ell\phi_{i}}v_{\ell,n}(q_{i})/\sqrt{2\pi}$ and $u_{-\ell,n}({\bf q}_{s})=e^{-i\ell\phi_{s}}v_{-\ell,n}(q_{s})/\sqrt{2\pi}$ represent the idler and signal Schmidt eigenmodes. These eigenmodes belong to the Laguerre-Gaussian basis. It is worth mentioning that although the Schmidt width is unique, other decompositions (actually infinitely many) are possible in different eigenmode bases. For instance, the Hermite-Gaussian modes (in Cartesian variables and the same $w_{S}=\sqrt{2w_{0}b}$) constitute another possible basis. In fact, all eigenmode bases connected via the following unitary (metaplectic) transformation $\hat{U}(\theta, \varphi) =\exp(-i\theta\,\hat{\boldsymbol{\mathcal{L}}}\cdot{\bf u}_{\varphi})$, where ${\bf u}_{\varphi}=(-\sin\varphi,\cos\varphi,0)$ is a unit vector in the equatorial plane of the so-called orbital Poincar\'{e} sphere (parameterised by the spherical angles $\theta$ and $\varphi$) and $\hat{\boldsymbol{\mathcal{L}}}$ is an angular momentum operator, form Schmidt bases for~(\ref{eq:SchmidtPhi}). The components of the angular momentum operator: $\hat{\mathcal{L}}_{x}$, $\hat{\mathcal{L}}_{y}$, and $\hat{\mathcal{L}}_{z}$ are the three SU(2) generators. Of these, only $\hat{\mathcal{L}}_{z}$ represents real spatial rotations along the propagation of light, and it is thus the only component associated with the orbital angular momentum of light that can be measured in experiments~\cite{CalvoPRA}.
\par
\section*{APPENDIX B}

In this second Appendix we explicitly show how to calculate the radial functions $\mathcal{R}_{\ell}(s)$ given in Eq.~(\ref{eq:LGCoincidence}). For the chosen two-photon amplitude~(\ref{eq:TPA}) and the measurement scenario, these radial functions depend on the spatial overlap of the Schmidt and the fiber (fundamental Gaussian) modes at the planes where the azimuthal phase plates are located. We need to evaluate the integrals $\mathcal{R}_{\ell}=\sum_{n}\!\!\sqrt{\lambda_{\ell n}}\int r_{i}r_{s}\textrm{d}r_{i}\textrm{d}r_{s}v_{\ell_{A}=0,n_{A}=0}^{(w_{G})}(r_{i})v_{\ell_{B}=0,n_{B}=0}^{(w_{G})}(r_{s})$ $\times v_{\ell,n}^{(w_{S})}(r_{i})v_{-\ell,n}^{(w_{S})}(r_{s})$. We use the Schmidt eigenvalues $\lambda_{\ell n}=(1-\xi^{2})^{2}\xi^{2\vert\ell\vert+4n}$ found in Appendix A. The idler and signal Schmidt modes $v_{\ell,n}^{(w_{S})}(r_{i})$ and $v_{-\ell,n}^{(w_{S})}(r_{s})$ belong to the Laguerre-Gaussian mode basis. We stress once again the fact that the corresponding fiber $w_{G}$ and Schmidt $w_{S}$ widths are not necessarily equal. This implies that the orthogonality relation for these modes, $\int r \textrm{d}r\, v_{\ell_{G},n_{G}}^{(w_{G})}(r)v_{\ell_{S},n_{S}}^{(w_{S})}(r)=\delta_{\ell_{G}\ell_{S}}\delta_{n_{G}n_{S}}$, does not hold in general when $w_{S}\neq w_{G}$. 
\par
We start by recalling a remarkable identity between Laguerre polynomials $L_{n}^{\vert\ell\vert}$ and the modified Bessel functions $I_{\ell}$~\cite{Gradshteyn}
\begin{eqnarray}
\sum_{n=0}^{\infty}\frac{n!L_{n}^{\vert\ell\vert}(x)L_{n}^{\vert\ell\vert}(y)}{\Gamma(\vert\ell\vert+n+1)}z^{n}&=&\frac{(xyz)^{-\vert\ell\vert/2}}{1-z}\exp\left[-\frac{(x+y)z}{1-z}\right]\nonumber\\
&\times&I_{\ell}\left(\frac{2\sqrt{xyz}}{1-z}\right) ,\label{eq:IdentityLGI}
\end{eqnarray}
valid for all real $x,y$, integer $\ell$ and complex $z$ ($\vert z\vert<1$). 
\par
In detail, the integrals to calculate read
\begin{eqnarray}
\mathcal{R}_{\ell}\!&=&\! (1-\xi^{2})\xi^{\vert\ell\vert}\sum_{n=0}^{\infty}\xi^{2n}\!\int_{0}^{\infty} r_{i}\textrm{d}r_{i}\frac{2\,e^{-\frac{r_{i}^{2}}{w_{G}^{2}}}}{w_{G}}\left[\frac{n!}{(\vert\ell\vert+n)!}\right]^{\frac{1}{2}}\nonumber\\
\!&\times&\!\left[\frac{\sqrt{2}r_{i}}{w_{S}}\right]^{\vert\ell\vert}L_{n}^{\vert\ell\vert}\left(\frac{2r_{i}^{2}}{w_{S}^{2}}\right)\frac{2e^{-\frac{r_{i}^{2}}{w_{S}^{2}}}}{w_{S}}\int_{0}^{\infty} r_{s}\textrm{d}r_{s}\frac{2e^{-\frac{r_{s}^{2}}{w_{G}^{2}}}}{w_{G}}\nonumber\\
\!&\times&\! \left[\frac{n!}{(\vert\ell\vert+n)!}\right]^{\frac{1}{2}}\left[\frac{\sqrt{2}r_{s}}{w_{S}}\right]^{\vert\ell\vert}L_{n}^{\vert\ell\vert}\left(\frac{2r_{s}^{2}}{w_{S}^{2}}\right)\frac{2e^{-\frac{r_{s}^{2}}{w_{S}^{2}}}}{w_{S}}\, .\label{eq:Rl1}
\end{eqnarray}
Inserting identity~(\ref{eq:IdentityLGI}) in~(\ref{eq:Rl1}) and employing the following result~\cite{Gradshteyn}
\begin{eqnarray}
\int_{0}^{\infty}\!\! r\,e^{-\gamma r^{2}} I_{\ell}(2\nu r)\,\textrm{d}r=\frac{\Gamma(1+\frac{\vert\ell\vert}{2})e^{\frac{\nu^{2}}{2\gamma}}}{2\Gamma(1+\vert\ell\vert)\nu\sqrt{\gamma}}M_{-\frac{1}{2},\frac{\vert\ell\vert}{2}}\!\left(\frac{\nu^{2}}{\gamma}\right)\!,\nonumber
\end{eqnarray}
valid for $\textrm{Re}(\gamma)>0$ and all complex $\nu$, with $M_{a,b}$ representing the Whittaker functions, we obtain
\begin{eqnarray}
\mathcal{R}_{\ell}\!&=&\!\frac{\Gamma(1+\frac{\vert\ell\vert}{2})}{\Gamma(1+\vert\ell\vert)}\left[\frac{8(1-\xi^{2})^{3}sw_{S}^{2}}{\xi^{3}w_{G}^{4}} \right]^{\frac{1}{2}}\nonumber\\
\!&\times&\! \int_{0}^{\infty}\!\textrm{d}r\exp\left\{-\frac{(2-s^{2})\xi r^{2}}{(1-\xi^{2})sw_{S}^{2}}\right\}\nonumber\\
\!&\times&\! M_{-\frac{1}{2},\frac{\vert\ell\vert}{2}}\!\left(\frac{2s\xi r^{2}}{(1-\xi^{2})w_{S}^{2}}\right) , \label{eq:Rl2}
\end{eqnarray}
where we have introduced the parameter $s=2\xi/[1+\xi^{2}+(1-\xi^{2})(w_{S}/w_{G})^{2}]$. The remaining integral is carried out through the use of formula
\begin{eqnarray}
\int_{0}^{\infty}\! \textrm{d}t\, \frac{e^{-\bar{\gamma} t}}{\sqrt{t}} M_{-\frac{1}{2},\frac{\vert\ell\vert}{2}}(t) \!&=&\!\frac{\Gamma(1+\frac{\vert\ell\vert}{2})}{\left(\bar{\gamma}-\frac{1}{2}\right)\left(\frac{1}{2}+\bar{\gamma}\right)^{\frac{\vert\ell\vert}{2}}}\nonumber\\
\!&\times&\!F\left(\frac{\vert\ell\vert}{2},\frac{\vert\ell\vert}{2};1+\vert\ell\vert;\frac{1}{\frac{1}{2}+\bar{\gamma}}\right)\!,\nonumber
\end{eqnarray}
valid when $\bar{\gamma}>1/2$. The function $F(a,b;c;d)$ denotes the hypergeometric function. Equation~(\ref{eq:Rl2}) is finally given by
\begin{eqnarray}
\mathcal{R}_{\ell}\!&=&\!\frac{1-\xi^{2}}{1+\frac{(1-\xi^{2})}{4}\left(\frac{w_{S}}{w_{G}}-\frac{w_{G}}{w_{S}}\right)^{2}}\frac{\Gamma^{2}(1+\frac{\vert\ell\vert}{2})}{\Gamma(1+\vert\ell\vert)}\nonumber\\
\!&\times&\! F\left(\frac{\vert\ell\vert}{2},\frac{\vert\ell\vert}{2};1+\vert\ell\vert;s^{2}\right)s^{\vert\ell\vert}. \label{eq:Rl3}
\end{eqnarray}
This expression reduces to Eq.~(\ref{eq:LGCoincidence}) when one leaves aside all the $\ell$-independent factors (they do not play any significant role in the normalised coincidence profiles). In this way the two-photon detection probabilities~(\ref{eq:Coincidence}) can be cast in terms of the impulse response functions $h^{(i)}_{\ell_{A},\ell}$, $h^{(s)}_{\ell_{B},-\ell}$ and the radial functions $\mathcal{R}_{\ell}(s)$ that solely depend on the parameter $s$ (when $\ell_{A}=\ell_{B}=0$). 
\par
\section*{APPENDIX C}
In this appendix we would like to justify the model proposed here, the double gaussian profile as phase matching function.
\subsection{Classical Energy for Non-linear Media}
 \label{sec:ClassicalEnergy}
In this section we derive explicitly the energy of an anisotropic and non-linear medium. This classical description enables us to find a reasonable quantized Hamiltonian which allow us to describe the PDC process. In anisotropic and non-linear media we must be cautious in the description of physical variables. For a deep analysis, we begin by assuming that the polarization of the material obeys the usual expansion from non-linear optics
 \begin{eqnarray}
          \nonumber
          P_{i}= \varepsilon_{o}[\chi_{ij}^{(1)} E_{j} + \chi_{ijk}^{(2)} E_{j}E_{k}
          + \chi_{ijkl}^{(3)} E_{j} E_{k} E_{l} + ...] \; , \\ \label{PDCPolarization}
 \end{eqnarray}
 where $\chi^{(n)}$ (with $n>1$) represent the nonlinear susceptibility tensors of order $n+1$ responsible for the coupling of $n+1$ fields. 
The energy of the electromagnetic field can be written as
 \begin{eqnarray}
          \label{PDCEnergy}
          W_{E}= \int d^{3}r \int_{o}^{D} {\bf E} \cdot d{\bf D} \; ,
 \end{eqnarray}
where the displacement vector field is defined as usual ${\bf D}= \epsilon_{o} {\bf E} + {\bf P}$. In linear media, the energy density of the electromagnetic field depends quadratically on the electric field $\propto E^{2}$, whereas in non-linear (parametric) media, polarization exhibits a dependence on higher order electric field terms (\ref{PDCPolarization}). We can expand the energy (\ref{PDCEnergy}) in two terms, one related to the linear contribution of the electric field and the other one related to the non-linear part. Taking the lowest order nonlinearity (\ref{PDCPolarization}), the energy due to this contribution can be written as \cite{Ghosh&Mandel,Hong&Mandel},
 \begin{eqnarray}
          \label{PDCInteractionHamiltonian}
          H_{I}= \frac{1}{2} \int_{\mathcal{V}} d^{3}r \; \widetilde{\chi}_{ijk}^{(2)} \; \fune{E_{i}}\fune{E_{j}}\fune{E_{k}}
 \end{eqnarray}
where the integration extends over the volume $\mathcal{V}$ of the nonlinear medium. Notice that the second order susceptibility is not the same as the susceptibility in equation (\ref{PDCPolarization}). We will consider that the medium response to the electric field is not instantaneous. Hence, the second order polarization should be written as
 \begin{eqnarray}
 	 \nonumber
          P_{i}^{(2)} ({\bf r},t) = \epsilon_{o} \int_{-\infty}^{\infty} \int_{-\infty}^{\infty} dt_{1} dt_{2} \; \chi_{ijk}^{(2)} (t-t_{1}, t-t_{2})  \\ \times \nonumber \; E_{j}({\bf r},t_{1}) E_{k}({\bf r},t_{2}) \; . \\ \label{eq:Polarization} 
 \end{eqnarray}
We can see, taking the Fourier transformation of the electric fields in equation (\ref{eq:Polarization}), that this specific form of the susceptibility implies that different monochromatic waves (with a well-defined frequency) of the electric field can excite other frequencies. Hence, the energy of the electromagnetic field due to the bilinear susceptibility (\ref{PDCInteractionHamiltonian}) can be expressed as \cite{Mandel&Wolf}, 
  \begin{eqnarray}
         \nonumber
          H_{I}= \frac{1}{2} \int_{\mathcal{V}} d^{3}r \int \int \int d\omega d\omega' d\omega'' \; \widetilde{\chi}_{ijk}^{(2)}(\omega, \omega', \omega'') \\  \times \nonumber \; E_{i}({\bf r},\omega'')  E_{j}({\bf r},\omega - \omega')  E_{k}({\bf r},\omega') \; . \\  \label{PDCInteractionHamiltonian2}
           \end{eqnarray}
This energy gives rise to the quantized Hamiltonian frequently found in the literature~\cite{Klyshko} to describe the process of spontaneous parametric down conversion, where an incident beam interacts with a non-linear medium and splits into two lower-frequency signal and idler photons. 


\subsection{Parametric Down-Conversion} 
\label{sec:PDC}

If we have a crystal with nonzero $\chi^{(2)}$, pumped with a
laser beam, there is a small, but non negligible, probability that a pump photon will decay into an idler and signal photon pair. We take the incident pump light beam to be in the form of an intense quasi-monochromatic plane wave propagating along the z-direction,
\begin{eqnarray}
\label{PumpBeam}
{\bf E}_{p} ({\bf r},t)= {\boldsymbol \epsilon}_{p} \; u_{nm} ({\bf r}_{\perp},z) \; e^{i\omega_{p} t}\; ,
\end{eqnarray}
where $ {\boldsymbol \epsilon}_{p}$ is the polarization of the pump beam and $u_{nm} ({\bf r}_{\perp},z)$ is a slowly-variant-amplitude (for the mode having indices $n$ and $m$) along $z$ which obeys the paraxial equation in anisotropic media \cite{GFC&Picon}. That is, the paraxial equation for free propagation cannot be used and must be modified. The pump beam is described classically, within the undepleted approximation, which means that one assumes that the input state is made up of a large number of photons and only few of them interact with the non-linear medium, producing pairs of correlated photons. In other words, the pump beam, at the output of the crystal, remains almost unaffected. \par
A form for the interaction Hamiltonian operator $\hat{H}_{I}(t)$ can be given motivated by the classical electromagnetic field energy (\ref{PDCInteractionHamiltonian2}), and reads (we are always in the interaction picture, though it can also be done in the Heisenberg picture \cite{Hong&Mandel}),
\begin{eqnarray}
\nonumber
\hat{H}_{I}(t) = \int_{\mathcal{V}} d^{3}r \; \frac{1}{L^{3}} \sum_{{\bf k_{i}}, \sigma_{i}} \sum_{{\bf k_{s}}, \sigma_{s}} \chi_{ijk}^{(2)} (\omega_{p},\omega_{i},\omega_{s}) \; ({\boldsymbol \epsilon}^{*}_{{\bf k_{i}}, \sigma_{i}})^{i} \\ \times \nonumber \; ({\boldsymbol \epsilon}^{*}_{{\bf k_{s}}, \sigma_{s}})^{j}  ({\boldsymbol \epsilon}_{p})^{k} \, u_{nm} ({\bf r}_{\perp},z) \,  \hat{a}^{\dagger}_{{\bf k_{i}}, \sigma_{i}} \hat{a}^{\dagger}_{{\bf k_{s}}, \sigma_{s}} \; e^{-i({\bf q}_{i}+{\bf q}_{s})\cdot {\bf r}_{\perp}} \;  \\ \times \; e^{-i\left[(k_{i z}+k_{s z})z + (\omega_{p}-\omega_{i}-\omega_{s})t\right]} + H. c.  \nonumber \; , \\ \label{GeneralInteractionHamiltonian}
\end{eqnarray}
where ${\bf k}^{(i,s)}={\bf q}^{(i,s)} + k_{z}^{(i,s)} \hat{{\bf u}}_{z}$. In equation (\ref{GeneralInteractionHamiltonian}) two quantized electric fields are considered; they will be responsible for idler and signal photons, whereas the pump beam is associated with the classical field \cite{Ghosh&Mandel}. As we have mentioned before, we are carrying out the calculations under the interaction picture, where we know that a state evolves as
{\footnotesize
 \begin{eqnarray}
          \nonumber
          \bra{\Phi (t)} = \mathbb{T} \left\{
          e^{-\frac{i}{\hbar}\int_{0}^{t} d\tau \hat{H}_{I}(\tau)} \right\} \bra{\Phi_{o}} \simeq \; \bra{\Phi_{o}}
         -\frac{i}{\hbar}\int_{0}^{t} d\tau \; :\hat{H}_{I}(\tau): \, \bra{\Phi_{o}}  \\ + \nonumber \left( \frac{i}{\hbar} \right)^{2} \mathbb{T} \left\{ \int_{0}^{t} d\tau \int_{0}^{t} d\tau' \; \hat{H}_{I}(\tau) \hat{H}_{I}(\tau') \right\} \, \bra{\Phi_{o}} + ... \; ,
  \end{eqnarray}
  \begin{eqnarray}       
         \label{StateEvolvingIP}
 \end{eqnarray}
 }with $\mathbb{T}$ denoting temporal ordering. Since the interaction is assumed to be weak enough, it suffices to expand the exponential (\ref{StateEvolvingIP}) up to first order term. The action of the Hamiltonian operator on the evolution of the initial vacuum state yields the initial state plus a two-photon state (contributions due to higher order processes involve the generation of four, six,... photons, and are exceedingly small). Notice that applying the Wick theorem in the above expansion, temporal ordering becomes normal ordering up to first order. Therefore,
 { \small
 \begin{eqnarray}
          \nonumber
          \bra{\Phi (t)}_{e} = \mathbb{T} \left\{
          e^{-\frac{i}{\hbar}\int_{0}^{t} d\tau \hat{H}_{I}(\tau)} \right\} \bra{\Omega} \simeq \; \bra{\Omega} -\frac{i}{\hbar}  \frac{1}{L^{3}} \sum_{{\bf k_{i}}, \sigma_{i}} \sum_{{\bf k_{s}}, \sigma_{s}} \\ \nonumber \chi_{ijk}^{(2)} (\omega_{p},\omega_{i},\omega_{s}) \; ({\boldsymbol \epsilon}^{*}_{{\bf k_{i}}, \sigma_{i}})^{i} ({\boldsymbol \epsilon}^{*}_{{\bf k_{s}}, \sigma_{s}})^{j}  ({\boldsymbol \epsilon}_{p})^{k} \\ \nonumber \int_{0}^{L} dz \; e^{-i(k_{i z}+k_{s z})z} \int_{\mathcal{S}} d^{2}{\bf r}_{\perp} \; u_{nm} ({\bf r}_{\perp},z) \, e^{-i({\bf q}_{i}+{\bf q}_{s})\cdot {\bf r}_{\perp}} \\ e^{-i(\omega_{p}-\omega_{i}-\omega_{s})t/2} \; \frac{\sin{\left[(\omega_{p}-\omega_{i}-\omega_{s})t/2 \right]}}{(\omega_{p}-\omega_{i}-\omega_{s})/2} \, \bra{{\bf k_{i}}, \sigma_{i};{\bf k_{s}}, \sigma_{s}} \; ,\nonumber \\ \label{GeneralStatePDC}
 \end{eqnarray}
 }where we have performed the integration with respect to time $\tau$. Let $L$ denote the length of the crystal along $z$, whereas $\mathcal{S}$ corresponds to the transverse area (the volume of the crystal is $\mathcal{V}=L \mathcal{S}$). Equation (\ref{GeneralStatePDC}) is a general expression for any input paraxial beam. Assuming that the transverse area of the crystal $\mathcal{S}$ is much larger than the characteristic width of the input beam, the integral with respect to the transverse position becomes the transverse Fourier transform. Now taking into account that the paraxial wave is propagating in an uniaxially anisotropic crystal \cite{GFC&Picon}, this integral becomes 
 \begin{widetext}
 {\small
 \begin{eqnarray} \label{FourierParaxAmplPDC}
\int_{\mathcal{S}} d^{2}{\bf r}_{\perp} \; u_{nm} ({\bf r}_{\perp},z) \, e^{-i({\bf q}_{i}+{\bf q}_{s})\cdot {\bf r}_{\perp}} \; \rightarrow \; e^{i k_{p} n_{e}(\theta,\omega_{p}) z} e^{- i \; \frac{n_{e}(\theta,\omega_{p})}{2n_{e}^{2}(\omega_{p})k_{p}} \left(q_{y}^{2} + q_{x}^{2}
        \frac{n_{e}^{2}(\theta,\omega_{p})}{n_{o}^{2}(\omega_{p})}\right)z} e^{-iq_{x} s_{\theta} c_{\theta} \;  \frac{n_{e}^{2}(\theta,\omega_{p})}{\tilde{n}^{2}(\omega_{p})}z} \; \widetilde{u}_{nm} ({\bf q}_{i}+{\bf q}_{s}) \, ,
\end{eqnarray}}
 \end{widetext}
 where, for simplicity, we assume that the optical axis is contained in the plane $xz$ and the pump beam is linearly polarized along the $x$-direction (the pump beam has extraordinary polarization). Here, $\theta$ refers to the angle between the optical axis and the z-axis, and
{\footnotesize
\begin{eqnarray*}
{\bf q} = {\bf q_{i}} + {\bf q_{s}} \hspace{8pt}Ê; \hspace{8pt} k_{p} = \frac{\omega_{p}}{c}  \hspace{8pt}Ê; \hspace{8pt} c_{\theta} \equiv \cos{\theta} \hspace{8pt} ; \hspace{8pt} s_{\theta} \equiv \sin{\theta} \hspace{8pt}Ê ,\\ \frac{1}{n_{e}^{2}(\theta, \omega_{p})} = \frac{\sin^{2}{\theta}}{n_{e}^{2}(\omega_{p})} + \frac{\cos^{2}{\theta}}{n_{o}^{2}(\omega_{p})}\hspace{8pt}Ê; \hspace{8pt} \frac{1}{\tilde{n}^{2}(\omega_{p})} = \frac{1}{n_{o}^{2}(\omega_{p})} - \frac{1}{n_{e}^{2}(\omega_{p})} \hspace{8pt}Ê.
\end{eqnarray*}
}Knowledge of the pump spectrum at the input crystal face enables, via equation (\ref{FourierParaxAmplPDC}), to describe the evolution of the pump beam through the crystal. In order to satisfy both the undepleted approximation and the fact that the pump profile is assumed not to vary significantly during the nonlinear propagation, the crystal must be very thin. The $z$-dependence in equation (\ref{FourierParaxAmplPDC}) will play a crucial role to correctly describe the phase-matching conditions below.  
\par
Defining
{\footnotesize
\begin{equation*}
\Theta \equiv \frac{1}{2 k_{p}} \frac{n_{e}(\theta,\omega_{p})}{n_{e}^{2}(\omega_{p})} \left(q_{y}^{2} + q_{x}^{2} \;
        \frac{n_{e}^{2}(\theta,\omega_{p})}{n_{o}^{2}(\omega_{p})}\right) - q_{x} \, \sin{\theta} \cos{\theta} \;  \frac{n_{e}^{2}(\theta,\omega_{p})}{\tilde{n}^{2}(\omega_{p})}\; ,
\end{equation*}
}we can perform the integral with respect to $z$ in equation (\ref{GeneralStatePDC}), resulting in the so-called phase-matching function:
{\small
\begin{eqnarray} 
\int_{0}^{L} dz \; e^{i[k_{p}n_{e} (\theta,\omega_{p}) - \Theta -k_{i z} -k_{s z}]z} = \nonumber \\ e^{i \Delta (k_{p}, {\bf k}_{i}, {\bf k}_{s}) L/2} \; \frac{\sin{\left[\Delta (k_{p}, {\bf k}_{i}, {\bf k}_{s}) L/2 \right]}}{\Delta (k_{p}, {\bf k}_{i}, {\bf k}_{s}) /2} \; ,\label{Mismatching}
\end{eqnarray}
}where the argument
\begin{eqnarray} \label{ArgMismatching}
\Delta (k_{p}, {\bf k}_{i}, {\bf k}_{s}) \equiv k_{p}n_{e}(\theta,\omega_{p}) -\Theta -k_{i z} -k_{s z} \; ,
\end{eqnarray}
describes the wave vector mismatching. Therefore, equation (\ref{GeneralStatePDC}) can be cast (excluding the vacuum state) as
\begin{widetext}
 \begin{eqnarray}
          \nonumber
          \bra{\Phi (t)}_{e}^{(2ph)} &=& -\frac{i}{\hbar}  \frac{1}{L^{3}} \sum_{{\bf k_{i}}, \sigma_{i}} \sum_{{\bf k_{s}}, \sigma_{s}} \chi_{ijk}^{(2)} (\omega_{p},\omega_{i},\omega_{s}) \; ({\boldsymbol \epsilon}^{*}_{{\bf k_{i}}, \sigma_{i}})^{i} ({\boldsymbol \epsilon}^{*}_{{\bf k_{s}}, \sigma_{s}})^{j}  ({\boldsymbol \epsilon}_{p})^{k} \; e^{i \Delta (k_{p}, {\bf k}_{i}, {\bf k}_{s}) L/2} \; \frac{\sin{\left[\Delta (k_{p}, {\bf k}_{i}, {\bf k}_{s}) L/2 \right]}}{\Delta (k_{p}, {\bf k}_{i}, {\bf k}_{s}) /2} \\ 
          &\times&\widetilde{u}_{nm} ({\bf q}_{i}+{\bf q}_{s}) \; e^{-i(\omega_{p}-\omega_{i}-\omega_{s})t/2} \; \frac{\sin{\left[(\omega_{p}-\omega_{i}-\omega_{s})t/2 \right]}}{(\omega_{p}-\omega_{i}-\omega_{s})/2} \, \; \bra{{\bf k_{i}}, \sigma_{i};{\bf k_{s}}, \sigma_{s}}\, .  \label{GeneralStatePDC2}
 \end{eqnarray}
 \end{widetext}
 
 \subsection{PDC in type II}
 
In this section we are going to consider the parametric down conversion under type II phase-matching conditions. This implies that the pump beam is extraordinarily (linearly) polarized, whereas the idler and idler photons are ordinarily (the polarization vector is perpendicular to the optical axis) and extraordinarily polarized, respectively. Our first aim will be to further simplify the down-converted state~(\ref{GeneralStatePDC2}). If the interaction time is sufficiently large, we can replace the temporal-sinc factor in equation (\ref{GeneralStatePDC2}) by a delta function,
\begin{eqnarray*}
\frac{\sin{\left[(\omega_{p}-\omega_{i}-\omega_{s})t/2 \right]}}{(\omega_{p}-\omega_{i}-\omega_{s})/2} \; \rightarrow \; 2 \pi \; \delta(\omega_{p}-\omega_{i}-\omega_{s}) \; ,
\end{eqnarray*}
This is a good approximation, and the oscillatory factor $e^{-i(\omega_{p}-\omega_{i}-\omega_{s})t/2}$ can then be discarded by a regularizing procedure. In particular, one obtains the frequency matching condition
\begin{eqnarray} \label{FreqMatchingCondition}
\omega_{p} = \omega_{i} + \omega_{s} \; ,
\end{eqnarray}
that is, energy conservation imposes that the annihilation of the pump photon gives rise to the creation of the idler and signal photons. In what follows, and effective nonlinear susceptibility $\chi_{eff}$ will be used to represent the relevant components of $\chi_{ijk}^{(2)}$ contracted with the corresponding components of the pump, idler and signal polarizations (we also absorb in $\chi_{eff}$ additional constants). Considering the continuity of the wave vectors of signal and idler photons (assuming that the volume of the crystal is large enough), equation (\ref{GeneralStatePDC2}) reduces to
{ \small
 \begin{eqnarray}
          \nonumber
          \bra{\Phi}_{e}^{(2ph)} = -\frac{i}{\hbar}  \int d^{3}{\bf k}_{i} \int d^{3}{\bf k}_{s} \; \chi_{eff} (\omega_{p},\omega_{i},\omega_{s}) \; e^{i \Delta (k_{p}, {\bf k}_{i}, {\bf k}_{s}) L/2} \\ \nonumber \times \; \frac{\sin{\left[\Delta (k_{p}, {\bf k}_{i}, {\bf k}_{s}) L/2 \right]}}{\Delta (k_{p}, {\bf k}_{i}, {\bf k}_{s}) L/2} \widetilde{u}_{nm} ({\bf q}_{i}+{\bf q}_{s}) \\ \nonumber \times \; \delta(\omega_{p}-\omega_{i}-\omega_{s}) \, \; \bra{{\bf k_{i}}, \sigma_{o},{\bf k_{s}}, \sigma_{e}}   \label{GeneralStatePDCtypeII} \; .
 \end{eqnarray}
 }Notice that the idler and signal photons have fixed polarization, ordinary ($\sigma_{o}$) and extraordinary ($\sigma_{e}$) respectively. 
\par
For convenience, we change the integral in the $z$ component of the wave vector by the frequency. Therefore, for the idler integral, the Jacobian due to the change of variables is
{\small
 \begin{eqnarray*}
d^{2}{\bf q}_{i}dk_{iz} = \left| \frac{\partial ({\bf q}_{i},k_{iz})}{\partial({\bf q}_{i},\omega_{i})} \right| d^{2}{\bf q}_{i}d\omega_{i} = \left| \frac{\partial k_{iz}}{\partial \omega_{i}} \right| d^{2}{\bf q}_{i}d\omega_{i} = \\ \nonumber \left| \frac{\omega_{i}}{k_{iz}}\frac{n_{o}^{2}(\omega_{i})}{c^{2}}+\frac{\omega_{i}^{2} n_{o}(\omega_{i})}{c^{2}k_{iz}} \frac{\partial n_{o}(\omega_{i})}{\partial\omega_{i}}\right| d^{2}{\bf q}_{i}d\omega_{i} \; ,
\end{eqnarray*}
}while for the signal integral, we have \begin{widetext}
 \begin{eqnarray} \nonumber
d^{2}{\bf q}_{s}dk_{sz} = \left| \frac{\partial ({\bf q}_{s},k_{sz})}{\partial({\bf q}_{s},\omega_{s})} \right| d^{2}{\bf q}_{s}d\omega_{s} = \left| \frac{\partial k_{sz}}{\partial \omega_{s}} \right| d^{2}{\bf q}_{s}d\omega_{s} = \\ \left| \frac{ 2 \frac{\omega_{s}}{c^{2}} + \frac{2 k_{sz}^{2}}{n_{e}^{3}(\theta, \omega_{s})}\frac{\partial n_{e}(\theta,\omega_{s})}{\partial \omega_{s}} - k_{sz}q_{sx} \sin{2\theta} \frac{\partial}{\partial \omega_{s}} \left( \frac{1}{\tilde{n}^{2}(\omega_{s})} \right) - q_{sx}^{2} \, \frac{\partial}{\partial \omega_{s}} \left(\frac{1}{n_{e}'^{2}(\theta,\omega_{s})} \right) + 2\frac{q_{sy}^{2}}{n_{e}^{3}(\omega_{s})}\frac{\partial n_{e}(\omega_{s})}{\partial \omega_{s}}}   {2 \frac{k_{sz}}{n_{e}^{2}(\theta, \omega_{s})}+\frac{\sin{2\theta}}{\tilde{n}^{2}(\omega_{s})}q_{sx}}
 \right| d^{2}{\bf q}_{s}d\omega_{s} \; . \label{differentialKzvsW}
\end{eqnarray} 
\end{widetext}

{\em \bf Proof}: {\small
For ordinary waves the dispersion relation in an anisotropic medium is
\begin{eqnarray*}
\frac{\omega_{i}^{2}n_{o}^{2} (\omega_{i})}{c^{2}} = k_{iz}^{2}+ {\bf q}_{i}^{2} \; .
\end{eqnarray*}
Solving the quadratic equation for the variable $k_{iz}$
\begin{eqnarray*}
k_{iz}=\sqrt{\frac{\omega_{i}^{2}n_{o}^{2} (\omega_{i})}{c^{2}} - {\bf q}_{i}^{2}} \; .
\end{eqnarray*}
Notice that we only choose the positive solution of $k_{iz}$. Finally, taking the derivative with respect to $\omega_{i}$
\begin{eqnarray*}
\frac{\partial k_{iz}}{\partial \omega_{i}}  =  \frac{\omega_{i}}{k_{iz}}\frac{n_{o}^{2}(\omega_{i})}{c^{2}}+\frac{\omega_{i}^{2} n_{o}(\omega_{i})}{c^{2}k_{iz}} \frac{\partial n_{o}(\omega_{i})}{\partial\omega_{i}} \; .
\end{eqnarray*}
For extraordinary waves the dispersion relation in an anisotropic uniaxial medium is more complex:
{\small
\begin{eqnarray} \nonumber
\frac{\omega_{s}^{2}}{c^{2}} = \frac{k_{sz}^{2}}{n_{e}^{2}(\theta,\omega_{s})}+\left(q_{sx} \cos{\varphi} + q_{sy} \sin{\varphi} \right) \frac{k_{sz}\sin{2\theta}}{\tilde{n}^{2}(\omega_{s})} + \\  \frac{(q_{sx}\cos{\varphi}+q_{sy}\sin{\varphi})^{2}}{n_{e}'^{2}(\theta,\omega_{s})}+ \frac{(q_{sx}\sin{\varphi}-q_{sy}\cos{\varphi})^{2}}{n_{e}^{2}(\omega_{s})}\; ,  \label{dispersionrelatione}
\end{eqnarray}
}where $\{\theta, \varphi\}$ are the polar and the azimuthal angles respectively in spherical coordinates, with $\theta$ referring to the angle subtended by the optical axis and the $z$-axis. For simplicity, as we have mentioned before, we assume the optical axis to be contained in the plane $x$-$z$ ($\varphi=0$). We have to take into account the common definitions in anisotropic media
\begin{eqnarray*}
\frac{1}{n_{e}^{2}(\theta,\omega_{s})}= \frac{\sin^{2}{\theta}}{n_{e}^{2}(\omega_{s})}+\frac{\cos^{2}{\theta}}{n_{o}^{2}(\omega_{s})} \, ,\\  \frac{1}{n_{e}'^{2}(\theta,\omega_{s})}= \frac{\cos^{2}{\theta}}{n_{e}^{2}(\omega_{s})}+\frac{\sin^{2}{\theta}}{n_{o}^{2}(\omega_{s})} \, ,\\ \frac{1}{\tilde{n}^{2}(\omega_{p})} = \frac{1}{n_{o}^{2}(\omega_{p})} - \frac{1}{n_{e}^{2}(\omega_{p})} \; .
\end{eqnarray*}
In this case, performing the implicit derivative with respect to $\omega_{s}$ in the dispersion relation (\ref{dispersionrelatione}),
\begin{eqnarray} \nonumber
2\frac{\omega_{s}}{c^{2}} = \frac{2k_{sz}}{n_{e}^{2}(\theta,\omega_{s})} \frac{\partial k_{sz}}{\partial \omega_{s}} -  \frac{2k_{sz}^{2}}{n_{e}^{3}(\theta,\omega_{s})}\frac{\partial n_{e}(\theta,\omega_{s})}{\partial \omega_{s}} + \\ \nonumber q_{sx} \frac{\sin{2\theta}}{\tilde{n}^{2}(\omega_{s})} \frac{\partial k_{sz}}{\partial \omega_{s}} + k_{sz} q_{sx} \sin{2\theta} \, \frac{\partial}{\partial \omega_{s}} \left( \frac{1}{\tilde{n}^{2}(\omega_{s})} \right) + \, \\ q_{sx}^{2} \, \frac{\partial}{\partial \omega_{s}} \left(\frac{1}{n_{e}'^{2}(\theta,\omega_{s})} \right) - 2\frac{q_{sy}^{2}}{n_{e}^{3}(\omega_{s})}\frac{\partial n_{e}(\omega_{s})}{\partial \omega_{s}} \; . \label{dem:dispersionrelatione}
\end{eqnarray}
where
\begin{eqnarray*}
\frac{\partial}{\partial \omega_{s}} \left( \frac{1}{\tilde{n}^{2}(\omega_{s})} \right) = -2 \left(\frac{1}{n_{o}^{3}(\omega_{s})}\frac{\partial n_{o}(\omega_{s})}{\partial \omega_{s}} - \frac{1}{n_{e}^{3}(\omega_{s})} \frac{\partial n_{e}(\omega_{s})}{\partial \omega_{s}} \right) \; , \\
\frac{\partial}{\partial \omega_{s}} \left(\frac{1}{n_{e}'^{2}(\theta,\omega_{s})} \right) = -2 \left( \frac{\cos^{2}{\theta}}{n_{e}^{3}(\omega_{s})} \frac{\partial n_{e}(\omega_{s})}{\partial \omega_{s}} + \frac{\sin^{2}{\theta}}{n_{o}^{3}(\omega_{s})}\frac{\partial n_{o}(\omega_{s})}{\partial \omega_{s}} \right) \; .
\end{eqnarray*}

Solving the equation (\ref{dem:dispersionrelatione}) for $\frac{\partial k_{sz}}{\partial \omega_{s}}$ we find equation (\ref{differentialKzvsW}).
}
\\ \par
Substituting equations (\ref{differentialKzvsW}) into the output state of idler and signal photons (\ref{GeneralStatePDCtypeII}), we obtain a general expression for PDC in type II. So far, we have only assumed the undepleted approximation and that the transverse area of the crystal is much larger than the characteristic width of the pump beam.

 \subsection{PDC in type II, collinear regime}
 \label{subsec:typeII_collinear}

 We now focus on the collinear regime in PDC where we only consider those idler and signal photons which are produced nearly in the same direction as that of the pump beam (in our framework along the $z$-axis). In this particular case, the approximation $\frac{q^{(s,i)}}{k_{z}^{(s,i)}} \ll 1$ is justified. Therefore, equations (\ref{differentialKzvsW}) reduce to
  \begin{eqnarray*}
d^{2}{\bf q}_{i}dk_{iz} = \frac{n_{o}(\omega_{i})}{c} d^{2}{\bf q}_{i}d\omega_{i} \; ,
\end{eqnarray*}
for the idler, and
 \begin{eqnarray} \nonumber
d^{2}{\bf q}_{s}dk_{sz} = \frac{n_{e}(\theta,\omega_{s})}{c} d^{2}{\bf q}_{s}d\omega_{s} \; , \label{differentialKzvsWCollinear}
\end{eqnarray} 
for the signal, where we have also made the reasonable assumption that the variation of refractive indices with respect to frequency is small. Thus, Jacobians due to the change of variables are just refractive indices, which depend on frequency.  
\par
In the collinear regime more simplifications can be done. Let us explicitly write the argument of the phase matching function (\ref{ArgMismatching}) as 
{\small
\begin{eqnarray} \nonumber
\Delta (k_{p}, {\bf k}_{i}, {\bf k}_{s}) = k_{p} n_{e}(\theta,\omega_{p}) - k_{s} n_{e}(\theta,\omega_{s}) - k_{i} n_{o} (\omega_{i}) + \\ \nonumber q_{sx} \frac{\sin{2\theta}}{2} \left[ \frac{n_{e}^{2}(\theta,\omega_{p})}{\tilde{n}^{2}(\omega_{p})}+ \frac{n_{e}^{2}(\theta,\omega_{s})}{\tilde{n}^{2}(\omega_{s})} \right] + q_{ix} \frac{\sin{2\theta}}{2} \frac{n_{e}^{2}(\theta,\omega_{p})}{\tilde{n}^{2}(\omega_{p})} - \\ \nonumber  \frac{n_{e}(\theta,\omega_{s})}{2n_{e}^{2}(\omega_{p})k_{p}} \left[ (q_{iy}+q_{sy})^{2} + (q_{ix}+q_{sx})^{2} \frac{n_{e}^{2}(\theta,\omega_{p})}{n_{o}^{2}(\omega_{p})} \right] +\\ \nonumber \frac{n_{e}^{3}(\theta,\omega_{s})}{2n_{e}^{2}(\omega_{s}) n_{o}^{2}(\omega_{s}) k_{s}} q_{sx}^{2} + \frac{n_{e}(\theta,\omega_{s})}{2n_{e}^{2}(\omega_{s})k_{s}} q_{sy}^{2}+ \frac{q_{ix}^{2}+q_{iy}^{2}}{2n_{o}(\omega_{i})k_{i}} \; .\\ \label{ArgMismatchingCollinear}
\end{eqnarray}
}It is clearly an strongly anisotropic function. Notice that the linear terms in the transverse wave vectors are responsible for the walk-off effect (the Poynting and wave vectors of the signal and pump photons are not collinear, i.e., their energy flows in a different direction from its corresponding phase front). We should mention a previous study by Torres and coworkers~\cite{Torres05} where the walk-off effect was taken into account to describe entangled photon states generated in type I PDC. 
\par
The state for idler and signal photons within type II PDC (\ref{GeneralStatePDCtypeII}) is
{\small
 \begin{eqnarray*}
          \bra{\Phi}_{e}^{(2ph)} =\int_{\Delta \Omega} \int_{0}^{\infty} d^{2}{\bf q}_{i}d\omega_{i} \int_{\Delta \Omega} \int_{0}^{\infty} d^{2}{\bf q}_{s}d\omega_{s} \; \chi_{eff}(\omega_{p},\omega_{i},\omega_{s}) \\ \times \nonumber \; e^{i \Delta (k_{p}, {\bf k}_{i}, {\bf k}_{s}) L/2} \; \textrm{sinc} \left[ \Delta (k_{p}, {\bf k}_{i}, {\bf k}_{s}) L/2 \right] \; \widetilde{u}_{nm} ({\bf q}_{i}+{\bf q}_{s}) \\ \times \; \delta(\omega_{p}-\omega_{i}-\omega_{s})  \, \; \bra{{\bf k_{i}}, \sigma_{o},{\bf k_{s}}, \sigma_{e}}   \;
 \end{eqnarray*}
 \begin{eqnarray}
\label{StatePDCtypeIICollinear}
\end{eqnarray}
}where $\textrm{sinc}(x) \equiv \sin(x)/x$. The symbol $\Delta\Omega$ represents the fact that the transverse idler and signal wave vectors are constrained to be very small (this is the condition for the collinear regime). This can be achieved experimentally by using a suitable pin hole placed in the $z$ axis. Now, due to this restriction, only the maximum of the sinc function contributes to the integral (the side lobes being blocked). Furthermore, the action of the pin hole truncating the sinc function can actually be modeled by a simple exponential function. In addition, we can regularize the phase factor $e^{i \frac{L}{4k_{p}} \vert {\bf q_{i}}-{\bf q_{s}} \vert^{2}}$ by just shifting our coordinate system ($z=0$) to the middle of the crystal, see equation (\ref{ArgMismatching}). We will also focus, for simplicity, on the frequency degenerated case, where the idler and signal photons have the same frequency ($\omega_{p}=\omega_{i}/2=\omega_{s}/2$, which can be realized by using appropriate narrow-band filters in front of the photodetectors). Thus, the two-photon state reads 
 \begin{eqnarray} \nonumber
          \bra{\Phi}_{e}^{(2ph)} = \chi_{eff}(\omega_{p}) \int d^{2}{\bf q}_{i} \int d^{2}{\bf q}_{s} \; e^{-\Delta (k_{p}, {\bf k}_{i}, {\bf k}_{s}) L/2} \\ \times \;  \widetilde{u}_{nm} ({\bf q}_{i}+{\bf q}_{s})  \, \; \bra{{\bf k_{i}}, \sigma_{o},{\bf k_{s}}, \sigma_{e}}   \label{StatePDCtypeIICollinear1} \; . \nonumber \\
 \end{eqnarray}
 The effective susceptibility, due to the energy conservation and the frequency degeneration in idler and signal photons, now depends only on the frequency of the pump beam. 
 \par
The argument of the phase-matching function $G=e^{-\Delta (k_{p}, {\bf k}_{i}, {\bf k}_{s}) L/2}$ is still too complex. Since we wish to maintain our framework at a sufficiently simple level, and we will pursue a full analytical study, some further (and drastic) approximations are required: we take all refractive indices (ordinary and extraordinary) to be identical and independent of frequency. Hence, the argument of the phase matching function reduces to 
\begin{eqnarray} \label{ArgMismatchingCollinearD}
\Delta (k_{p}, {\bf k}_{i}, {\bf k}_{s}) = \frac{1}{2k_{p}n_{eff}} \vert {\bf q_{i}}-{\bf q_{s}} \vert^{2} \; ,
\end{eqnarray} 
which is, in contrast to equation (\ref{ArgMismatchingCollinear}), an isotropic (with cylindrical symmetry) function. The effective refractive index $n_{eff}$ can be interpreted as a suitable fitting parameter (in principle, it can be quite different from the ordinary and extraordinary refractive indices).
\par
\begin{figure}[b]
\begin{center}
\hspace*{-0.15cm}
\hbox{\vbox{\vskip 0.0cm \includegraphics[width=60mm]{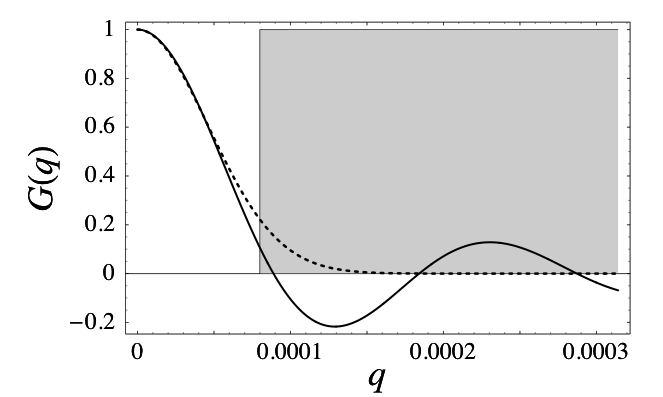}}}
\end{center}
\vspace*{-0.6cm}
\caption{\small Comparison between the sinc (solid curve) and the exponential (dotted curve) phase-matching functions when their arguments are given by equations (\ref{ArgMismatchingCollinear}) and (\ref{ArgMismatchingCollinearD}), respectively. The shadowed region corresponds to the truncation introduced by the pin hole.}
\label{fig:Check}
\end{figure}

In spite of the strong approximations performed above, it is illustrative to determine their goodness by comparing the sinc and the exponential  phase-matching functions when their arguments are given by equations (\ref{ArgMismatchingCollinear}) and (\ref{ArgMismatchingCollinearD}), respectively. Figure \ref{fig:Check} depicts both profiles. It can be seen that for sufficiently small transverse wave vectors both profiles are coincident (by suitably choosing $n_{eff}$); the deviations arise, typically, when the pin hole is already truncating both functions. This qualitatively justifies the validity of the approximations made above. Actually, this approximation was previously used (without any justification) by Law and Eberly~\cite{Law04} to study bipartite transverse entanglement from type II PDC. Furthermore, in other previous works, the sinc phase matching function was even replaced by a constant~\cite{Aiello04,Aiello05,Franke,Torres03}.
\par
\par
In summary, equation (\ref{StatePDCtypeIICollinear1}) finally becomes (simplifying the notation and absorbing the constant factors)
  { \small
 \begin{eqnarray}
          \bra{\psi} =\int d^{2}{\bf q}_{i} \int d^{2}{\bf q}_{s} \;\Phi({\bf q}_{i},{\bf q}_{s})  \, \; \bra{{\bf q}_{i},{\bf q}_{s}}   \label{StatePDCtypeIICollinear2} \; ,
 \end{eqnarray}
 }where $\Phi=EG$ is the two-photon amplitude, with $E({\bf q}_{i}+{\bf q}_{s})=\widetilde{u}_{nm}({\bf q}_{i}+{\bf q}_{s})$ denoting the transverse (spectrum) profile of the pump beam, and $G({\bf q}_{i}-{\bf q}_{s})=e^{- \frac{L}{4k_{p}n_{eff}} \vert {\bf q_{i}}-{\bf q_{s}} \vert^{2}}$ being the phase-matching function. Since we will not focus in later sections on the polarization degree of freedom (recall that idler and signal photons are assumed to be linearly polarized), we have omitted its explicit dependence in (\ref{StatePDCtypeIICollinear2}).
\par 
Equation (\ref{StatePDCtypeIICollinear2}) will be exploited in (\ref{eq:TPA}) to carry out our discussion about the OAM entanglement shared by the downconverted idler and signal photons.

\end{document}